\documentclass[prl,aps,twocolumn,nofootinbib,preprintnumbers]{revtex4}
\usepackage{graphicx}

\begin{document}

\def\beq{\begin{eqnarray}}
\def\eeq{\end{eqnarray}}
\newcommand{\gsim}{ \mathop{}_{\textstyle \sim}^{\textstyle >} }
\newcommand{\lsim}{ \mathop{}_{\textstyle \sim}^{\textstyle <} }
\newcommand{\vev}[1]{ \left\langle {#1} \right\rangle }
\newcommand{\bra}[1]{ \langle {#1} | }
\newcommand{\ket}[1]{ | {#1} \rangle }
\newcommand{\EV}{ {\rm eV} }
\newcommand{\KEV}{ {\rm keV} }
\newcommand{\MEV}{ {\rm MeV} }
\newcommand{\GEV}{ {\rm GeV} }
\newcommand{\TEV}{ {\rm TeV} }
\newcommand{\bea}{\begin{eqnarray}}   
\newcommand{\eea}{\end{eqnarray}}
\newcommand{\bear}{\begin{array}}  
\newcommand {\eear}{\end{array}}
\newcommand{\bef}{\begin{figure}}  
\newcommand {\eef}{\end{figure}}
\newcommand{\bec}{\begin{center}}  
\newcommand {\eec}{\end{center}}
\newcommand{\non}{\nonumber}  
\newcommand {\eqn}[1]{\beq {#1}\eeq}
\newcommand{\la}{\left\langle}  
\newcommand{\ra}{\right\rangle}
\newcommand{\ds}{\displaystyle}
\def\SEC#1{Sec.~\ref{#1}}
\def\FIG#1{Fig.~\ref{#1}}
\def\EQ#1{Eq.~(\ref{#1})}
\def\EQS#1{Eqs.~(\ref{#1})}
\def\GEV#1{10^{#1}{\rm\,GeV}}
\def\MEV#1{10^{#1}{\rm\,MeV}}
\def\KEV#1{10^{#1}{\rm\,keV}}
\def\lrf#1#2{ \left(\frac{#1}{#2}\right)}
\def\lrfp#1#2#3{ \left(\frac{#1}{#2} \right)^{#3}}

%


\preprint{UT-13-18~~TU-932~~IPMU13-0070}
\title{Polynomial Chaotic Inflation in the Planck Era}
\renewcommand{\thefootnote}{\alph{footnote}}

\author{Kazunori Nakayama$^{a,c}$,
Fuminobu Takahashi$^{b,c}$
and 
Tsutomu T. Yanagida$^{c}$}

\affiliation{
 $^a$Department of Physics, University of Tokyo, Tokyo 113-0033, Japan\\
 $^b$Department of Physics, Tohoku University, Sendai 980-8578, Japan\\
 $^c$Kavli IPMU, TODIAS, University of Tokyo, Kashiwa 277-8583, Japan
  }

\begin{abstract}
\noindent
We propose a chaotic inflation model in supergravity based on  polynomial interactions of
the inflaton. Specifically we study the chaotic inflation model with quadratic, cubic and quartic
couplings in the scalar potential and show that the predicted scalar spectral index and tensor-to-scalar ratio can lie 
within the $1\sigma$ region allowed by the Planck results. 
\end{abstract}

\maketitle

There is overwhelming observational evidence for inflation in the early Universe~\cite{Guth:1980zm}.
Not only does the inflation provide beautiful explanations for the observed homogeneity and isotropy of the Universe,
but  it also predicts tiny density fluctuations  with distinct properties; namely, they are nearly scale-invariant and Gaussian.
Both properties have been confirmed with unprecedented accuracy by the Planck satellite~\cite{Ade:2013rta}.
The last missing piece toward the observational confirmation of the inflation theory will be detection of the
primordial gravitational waves generated during inflation. 
Although the strength of the gravitational waves  depends on inflation models,
a simple class of models, called chaotic inflation~\cite{Linde:1983gd}, predicts 
gravitational waves with a large amplitude within the reach of the Planck satellite and future CMB observation experiments.

The recent Planck observations have tightly constrained possible inflation models~\cite{Ade:2013rta}.
Specifically, a chaotic inflation model
based on a quartic potential is highly disfavored by the observation, and that 
based on a quadratic potential is marginally consistent with the observation
at $2 \sigma$ level. Those with a linear or fractional power potential lie
outside $1\sigma$ but within $2\sigma$ allowed region. When combined with the
polarization data, the Planck data will be able to discover or put even severe constraints
on those inflation models.

The inflaton field slowly rolls down to the minimum of its potential during inflation, and 
it travels more than the Planck scale 
in the chaotic inflation. In order to control the inflaton dynamics over such
large variation, it is worthwhile to build a successful chaotic inflation model
in a framework of supergravity or superstring theory.
In this letter we extend the chaotic inflation model in supergravity proposed by Kawasaki, Yamaguchi,
and one of the present authors (Yanagida)~\cite{Kawasaki:2000yn}, 
by allowing higher order interactions of the inflaton in the superpotential. 

We start with the following K\"ahler and super-potentials,\footnote{
This type of generalization was considered in Refs.~\cite{Kawasaki:2000ws, Kallosh:2010ug}.
In Ref.~\cite{Kawasaki:2000ws}, it was pointed out that the form of the superpotential, $W= X \phi^n$, 
leads to a chaotic inflation based on $\varphi^{2n}$ potential.  General inflation models based on $W= X f(\phi)$ 
were studied  in Refs.~\cite{Kallosh:2010ug,Kallosh:2010xz}.
See Refs.~\cite{Takahashi:2010ky,Nakayama:2010kt,Harigaya:2012pg} for the realization of inflation models based on a
linear or fractional power potential in supergravity and 
also Refs.~\cite{Silverstein:2008sg,McAllister:2008hb,Peiris:2013opa} in string theory.
}
\bea
\label{KW0}
K&=& c_1(\phi+\phi^\dag) + \frac{1}{2} (\phi+\phi^\dag)^2 + |X|^2+\cdots,\\
W &=&   X \left(d_0 + d_1 \phi + d_2 \phi^2 + d_3 \phi^3+\cdots \right),
\label{KW}
\eea
where $\phi$ and $X$ are chiral superfields, and $c_1$, and $d_i$ are real and complex numerical coefficients, 
and the dots represent higher-order interactions. In the following,
the first term in the superpotential, $d_0$, is taken to be zero by an appropriate shift of $\phi$.
Here and in what follows we adopt the Planck unit where
$M_P \simeq 2.4 \times 10^{18}$\,GeV is set to be unity. 
The K\"ahler potential respects a shift symmetry $\phi \to \phi + iC$ with a real constant $C$.
Thanks to the shift symmetry, the imaginary component of $\phi$, $\varphi \equiv \rm{Im}[\sqrt{2}\phi]$, 
 does not appear in the K\"ahler potential, which enables us to identify $\varphi$ with the inflaton.
 The real component, $\sigma \equiv \rm{Re}[\sqrt{2}\phi]$, on the other hand, receives supergravity
 corrections and cannot go beyond the Planck scale. 
We also assume that $X$ and $\phi$ have an $R$-charge $+2$ and $0$, respectively.
The superpotential terms involving $\phi$ explicitly break the shift symmetry, and hence $|d_1|, |d_2|, \cdots$ 
are much smaller than unity,  and can be viewed as
the order parameters of the shift symmetry breaking.

In our analysis we focus on the $d_1$ and $d_2$ terms in the superpotential, and drop the higher-order terms, 
assuming they are sufficiently suppressed.\footnote{This will be the case if 
the effective cut-off scale in the superpotential is more than one order of magnitude larger than the Planck scale.}
Let us define $m \equiv |d_1|$, $\lambda \equiv |d_2|$ and $\theta \equiv {\rm arg}[d_1^*d_2]$ for later use.
For the moment we take $c_1=0$, and the effect of non-zero $c_1$ will be discussed later.

\begin{figure}[t!!]
\begin{center}
\includegraphics[scale=1.2]{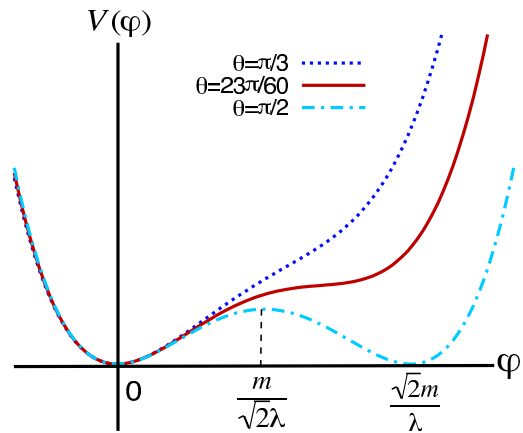}
\caption{ 
	The schematic picture of the scalar potential (\ref{scalar}).
}
\label{fig:pot}
\end{center}
\end{figure}

Substituting (\ref{KW0}) and (\ref{KW}) into the general expression for the scalar potential in supergravity
\begin{equation}
	V \;=\; e^K\left[ K^{i\bar j} (D_iW)(D_{\bar j}\bar W) -3|W|^2 \right],
\end{equation}
we find
\begin{equation}
	V \;\simeq\; \frac{1}{2}\varphi^2
	\left( m^2 - \sqrt{2} m \lambda \sin \theta\, \varphi + \frac{\lambda^2}{2} \varphi^2 \right).
	\label{scalar}
\end{equation}
Here we have  assumed that both $\sigma$ and $X$ are stabilized at $|\sigma| \ll 1$ and 
$\la X \ra \sim 0$ due to the higher order terms in the K\"ahler potential (\ref{KW0}).
Note that $\la W \ra$ is suppressed due to $\la X \ra \sim 0$, which enables the inflation
for $\varphi \gg 1$. 
The schematic picture of the scalar potential is shown in Fig.~\ref{fig:pot} for
a few different values of $\theta$. The quadratic chaotic inflation model is reproduced in the limit of $\lambda \rightarrow 0$.
For $\sin \theta > 0$, the second term in (\ref{scalar})
gives a negative contribution to the scalar potential, making the potential flatter or negatively curved
at large $\varphi$.
   It is worth stressing that a variety of the inflation models can 
be realized  simply by taking the $d_1$ and $d_2$  terms  in (\ref{KW}) with a different relative phase.
Note that this is effectively a single-field inflation, since the other degrees of freedom can be safely
stabilized during inflation.

Before proceeding further, let us mention the works in the past. In Ref.~\cite{Destri:2007pv},
the inflation model based on the scalar potential equivalent to (\ref{scalar}) was studied  
in a non-supersymmetric framework, where the inflaton is a real scalar field. Recently, the 
model was revisited and its global supersymmetric extension was proposed in Ref.~\cite{Croon:2013ana},
where it was shown that the predicted spectral index and the tensor-to-scalar ratio can be
consistent with the Planck data for $\theta \approx \pi/2$. However, the inflaton is a complex scalar field
and it is not clear how to stabilize the inflationary trajectory for $\theta \ne \pi/2$. 
We also note that, in supergravity, the large expectation of their $\la W \ra$ would lead to a negative scalar 
potential for the inflaton field value greater than the Planck scale, spoiling the inflation.
This is known as one of the difficulties to implement chaotic inflation in supergravity with a single superfield. 
The latter problem can be avoided in the no-scale supergravity~\cite{Cremmer:1983bf,Ellis:1983sf,Murayama:1993xu}.

Now let us continue with our discussion of the inflation model (\ref{scalar}).
The global shape of the potential depends on the relative phase $\theta$ 
as shown in Fig.~\ref{fig:pot}. In a case of $\theta = \pi/2$, 
there appear one local maximum at $\varphi = m/(\sqrt{2}\lambda) \equiv \varphi_t$
and two degenerate minima at $\varphi = 0$ and $2 \varphi_t$.\footnote{
In the case of $\theta = \pi/2$, the inflaton dynamics is equivalent to that in the spontaneous 
symmetry breaking model first considered in \cite{Linde:1983fq}. See also Ref.~\cite{Kallosh:2007wm}.
}
For smaller values of $\theta$, the minimum at $\varphi \ne 0$ is lifted.
As long as there are such local minimum and maximum, 
the initial  value of the inflaton field should be below the local maximum 
since otherwise the inflaton would be trapped in the false vacuum. 
The false vacuum disappears for $|\sin \theta| < 2\sqrt 2/3$. In this case 
successful chaotic inflation takes place for an arbitrary large initial field value~\cite{Linde:1983gd}.
Interestingly, if the relative phase $\theta$ marginally satisfies the inequality, there appears a flat
plateau at around $\varphi = \varphi_t$. As we shall see shortly, the predicted spectral index as well as
the tensor-to-scalar ratio are then significantly modified and they can lie within the $1\sigma$ allowed region.

\begin{figure}
\begin{center}
\includegraphics[scale=0.6]{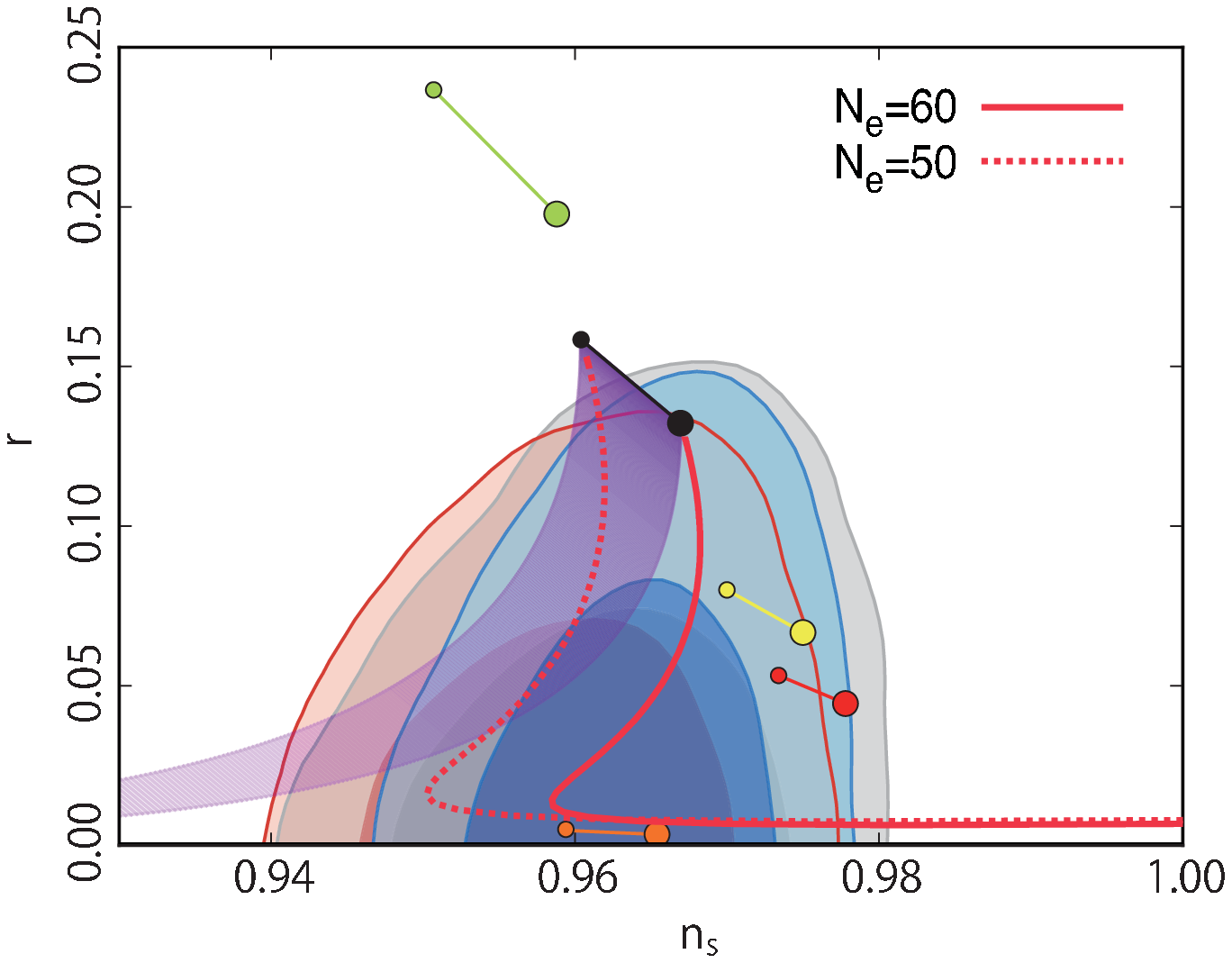}
\includegraphics[scale=0.6]{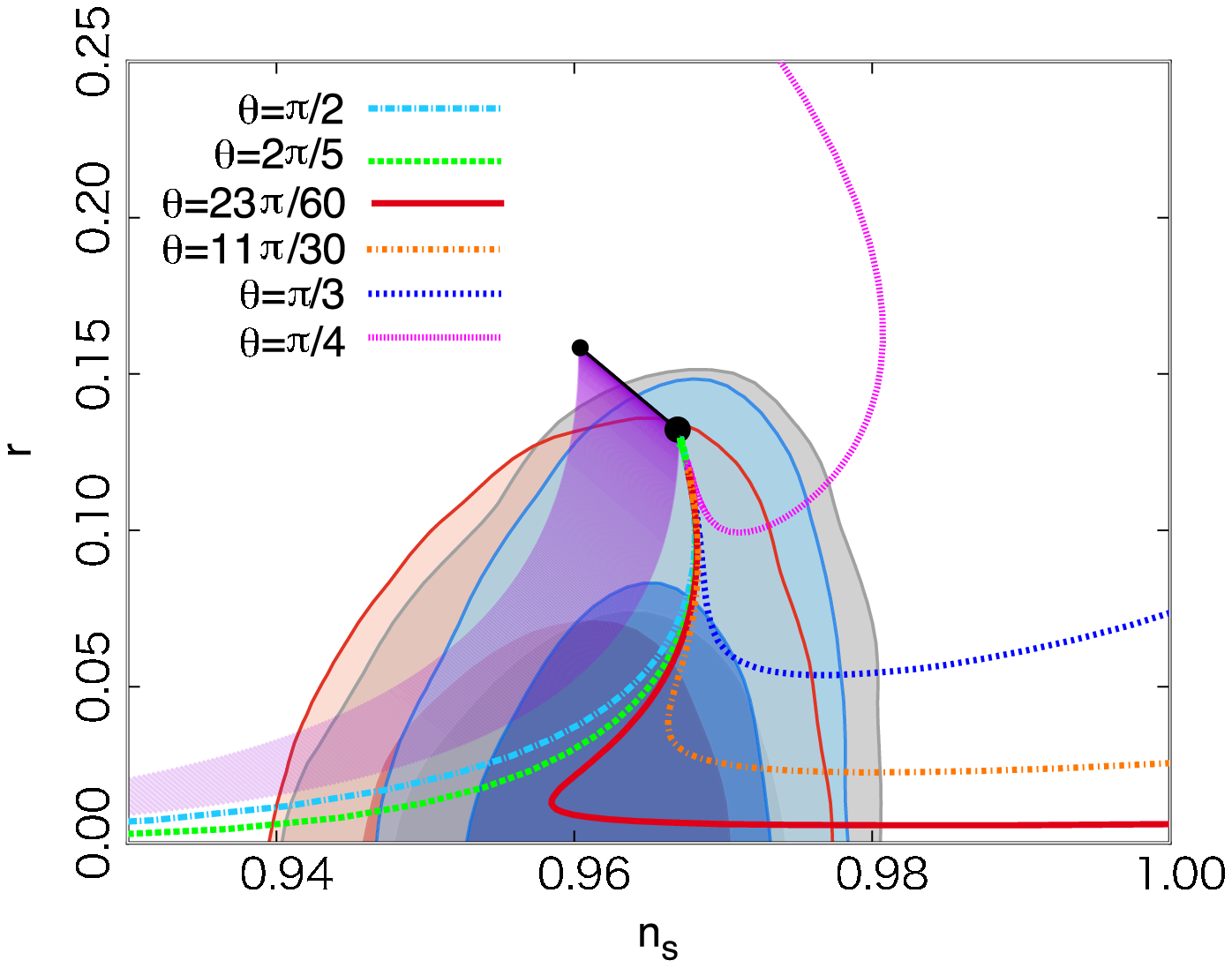}
\caption{ (Top) The prediction of polynomial chaotic inflation model is shown in $(n_s,r)$ plane for
$N_e = 60$ (red, solid) and $50$ (red, dashed). In this plot we have taken $\theta = 23\pi/60$.
Also shown are observational $1\sigma$ (dark) and $2\sigma$ (light) constraints from the Planck satellite~\cite{Ade:2013rta}: 
Planck + WMAP polarization (gray), Planck + WMAP polarization + high-$\ell$ CMB measurement (red),
Planck + WMAP polarization + baryon acoustic oscillation (blue).
Filled circles connected by line segments show the predictions from chaotic inflation with $V \propto \varphi^3$ (green), $\varphi^2$ (black), $\varphi$ (yellow), $\varphi^{2/3}$ (red) and $R^2$ inflation (orange),
for $N_e=50$ (small circle)--$60$ (big circle).
Purple band shows the prediction of natural inflation~\cite{Ade:2013rta}.
(Bottom) Same as top panel, but for various values of $\theta$. Here we have taken $N_e=60$.
}
\label{fig:ns-r}
\end{center}
\end{figure}

The chaotic inflation with quadratic potential, $V=m^2\varphi^2/2$,
is reproduced in the limit $\varphi_t \gg \mathcal O(10)$. As mentioned before,
the model is at odds with the recent Planck result at the 2$\sigma$ level.
An interesting situation appears when $\varphi_t \sim \mathcal O(10)$.
In this case, the last $50$ or $60$ e-foldings of the inflation occurs around $\varphi \sim \varphi_t$
where the potential is flatter due to the contribution from the $\varphi^3$ and $\varphi^4$ terms.
As a result, the inflation energy scale can be significantly lowered
compared with the quadratic chaotic inflation model,
when the Planck normalization of the density perturbations is imposed. 
Thus it will be able to relax the tension between the prediction of the 
chaotic inflation model and observations.

\begin{figure}[t!!]
\begin{center}
\includegraphics[scale=1.3]{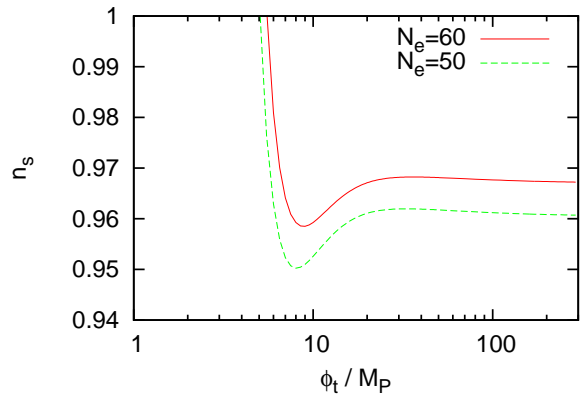}
\caption{ 
	The scalar spectral index as a function of $\varphi_t/M_P$ for $\theta = 23\pi/60$.
}
\label{fig:ns}
\end{center}
\end{figure}

We have numerically solved the equation of motion of $\varphi$ with the scalar potential $(\ref{scalar})$ 
and calculated the scalar spectral index, $n_s = 1-6\epsilon + 2\eta$, and the tensor-to-scalar ratio, $r=16\epsilon$, where
$2\epsilon = (V'/V)^2$ and $\eta = V^{''}/V$ evaluated at $\varphi = \varphi(N_e)$~\cite{Liddle:2000cg}.
Here $\varphi(N_e)$ is calculated from
\begin{equation}
	N_e = \int^{\varphi(N_e)}_{\varphi_{\rm end}} \frac{V}{V'}d\varphi,
\end{equation}
where $\varphi_{\rm end}$ denotes the field value at the end of inflation, at which ${\rm max}[ \epsilon, |\eta| ]=1$.
The results are shown in the top panel of Fig.~\ref{fig:ns-r} $N_e = 60$ (red, solid) and $50$ (red, dashed)
together with observational constraints from the Planck satellite~\cite{Ade:2013rta}.
Here we have taken $\theta = 23\pi/60$, which marginally satisfies the condition for the disappearance of the false vacuum leading to a flat plateau in the inflaton potential.
In the bottom panel of Fig.~\ref{fig:ns-r}, results for various values of $\theta$ are shown for $N_e=60$.
We can clearly see that almost entire region allowed by the Planck data can be covered by our model,
and importantly, the predicted value of $r$ is testable in future/on-going B-mode polarization search experiments.
Note that the predicted $(n_s, r)$ lies within the $1\sigma$ region for $\theta = \pi/3 \sim \pi/2$. 
In the following we take $\theta = 23\pi/60$ unless otherwise stated.
In Fig.~\ref{fig:ns}, the scalar spectral index as a function of $\varphi_t$ is shown.
It is seen that in the large $\varphi_t$ limit, the prediction approaches to that of the chaotic inflation with quadratic potential, as expected.
By choosing $\varphi_t = O(10)$, the predicted $n_s$ and $r$ can lie within the $1\sigma$ region allowed by the Planck data.
In particular, the predicted $r$ is testable in future/on-going B-mode polarization search experiments.

\begin{figure}[t!!]
\begin{center}
\includegraphics[scale=1.3]{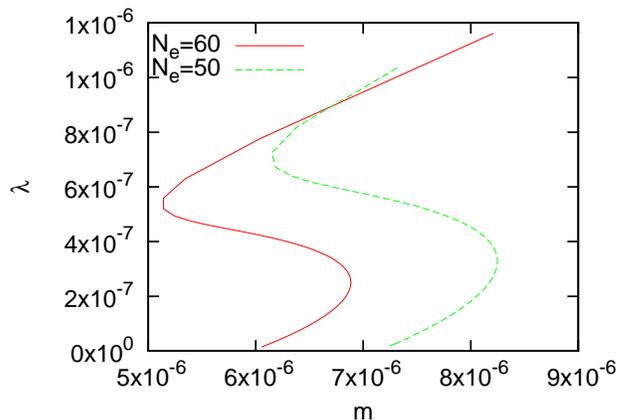}
\caption{ 
	The parameters $\lambda$ and $m$ (in Planck unit) which reproduce the
	Planck normalization of the CMB anisotropy for $\theta = 23\pi/60$.
}
\label{fig:lam}
\end{center}
\end{figure}

While the predicted $n_s$ and $r$ depends only on $m/\lambda$,  the Planck normalization of the
CMB anisotropy fixes a relation between $m$ and $\lambda$. We have confirmed that they are approximately given by $m \simeq (1 -2)\times \GEV{13}$ and $\lambda \lesssim 5 \times 10^{-7}$ 
for the spectral index allowed by the Planck data as shown in Fig.~\ref{fig:lam}. 
In terms of $d_1$ and $d_2$, they are roughly related to each other as $|d_2| = O(0.01) |d_1|$
for the parameters of our interest.

%
%

The reheating can be induced by introducing the following coupling to the Higgs doublets in the superpotential:
\begin{equation}
	W \supset \kappa XH_u H_d,
\end{equation}
with a numerical constant $\kappa$.
This allows the $\phi (X)$ decay into the Higgs boson and higgsino pair. Note here that 
$\phi$ and $X$ are maximally mixed with each other to form the mass eigenstate around the vacuum.
The reheating temperature is estimated as
\begin{equation}
	T_{\rm R} \sim 4\times 10^{9}\,{\rm GeV} \left( \frac{\kappa}{10^{-5}} \right)\left( \frac{m}{10^{13}\,{\rm GeV}} \right)^{1/2}.
\end{equation}
Therefore, thermal leptogenesis~\cite{Fukugita:1986hr} works, and
the Wino LSP produced by the decay of thermally produced gravitinos 
can account for the observed dark matter abundance in a heavy gravitino scenario.

Finally we mention the gravitino production from the inflaton decay~\cite{Kawasaki:2006gs,Endo:2006qk,Endo:2007ih,Endo:2007sz}.
Assuming that the dominant channel of the gravitino production is that through the inflaton decay into the hidden hadrons~\cite{Endo:2007ih}, the gravitino abundance is given by
\begin{equation}
	\frac{n_{3/2}}{s} \simeq 9\times 10^{-14}\left( \frac{10^{9}\,{\rm GeV}}{T_{\rm R}} \right)
	\left( \frac{\langle\phi\rangle}{10^{15}\,{\rm GeV}} \right)^{2}
	\left( \frac{m}{10^{13}\,{\rm GeV}} \right)^{2},
\end{equation}
where $n_{3/2}$ and $s$ are the gravitino number density and the entropy density, respectively.
Thus the gravitino abundance crucially depends on the VEV of the inflaton, and its impact on cosmology
depends on the gravitino mass. In general, $c_1$ in (\ref{KW0}) is non-zero and then the VEV is effectively 
given by $\langle\phi\rangle \sim c_1$. To be concrete, we assume a heavy gravitino, $m_{3/2} = O(10-100)$\,TeV.
Then we need $c_1 \lesssim10^{-3}$ to avoid the overproduction of the LSPs produced by the gravitino decay. 
Alternatively,  if the $R$-parity is violated by a small amount, the LSPs thus produced soon disappear
before the big-bang nucleosynthesis begins. 
There is no cosmological gravitino problem in such a case.\footnote{
	See also Ref.~\cite{Nakayama:2012hy} to suppress the gravitino overproduction from the inflaton decay.
}

In summary, we have proposed a polynomial chaotic inflation model defined in (\ref{KW0}) and (\ref{KW}).
Focusing on the $d_1$ and $d_2$ terms in the superpotential, we have shown that
it is possible to realize a wide variety of inflation models due to the different relative phase $\theta$.
Importantly, the inflation dynamics is described by single-field inflation, since the other 
degrees of freedom can be safely stabilized. 
We studied the inflaton dynamics for various values of $\theta$, and showed that
the predicted spectral index and tensor-to-scalar ratio can lie
within the $1\sigma$ region allowed by the Planck results for $\pi/3 \lesssim \theta \lesssim \pi/2$.
Interestingly, the tensor-to-scalar ratio is relatively large, and will be
 testable in future/on-going B-mode polarization searches.

\section*{Acknowledgments}

We are grateful to Andrei Linde for letting us know about Ref.~\cite{Destri:2007pv}
and his works~\cite{Linde:1983fq,Kallosh:2007wm}. TTY thanks John Ellis for useful
communication. 
This work was supported by the Grant-in-Aid for Scientific Research on
Innovative Areas (No. 21111006  [KN and FT],  No.23104008 [FT], No.24111702 [FT]),
Scientific Research (A) (No. 22244030 [KN and FT], 21244033 [FT], 22244021 [TTY]), and JSPS Grant-in-Aid for
Young Scientists (B) (No.24740135) [FT].  This work was also
supported by World Premier International Center Initiative (WPI Program), MEXT, Japan.




\begin{thebibliography}{99}


\bibitem{Guth:1980zm} 
  A.~H.~Guth,
  Phys.\ Rev.\ D {\bf 23}, 347 (1981);
  A.~A.~Starobinsky,
  Phys.\ Lett.\ B {\bf 91}, 99 (1980);
  K.~Sato,
  Mon.\ Not.\ Roy.\ Astron.\ Soc.\  {\bf 195}, 467 (1981).


\bibitem{Ade:2013rta} 
  P.~A.~R.~Ade {\it et al.}  [ Planck Collaboration],
  arXiv:1303.5082 [astro-ph.CO].
  
\bibitem{Linde:1983gd}
  A.~D.~Linde,
  Phys.\ Lett.\ B {\bf 129}, 177 (1983).
  



\bibitem{Kawasaki:2000yn} 
  M.~Kawasaki, M.~Yamaguchi, T.~Yanagida,
  Phys.\ Rev.\ Lett.\  {\bf 85}, 3572 (2000)
  [hep-ph/0004243].
  
\bibitem{Kawasaki:2000ws} 
M.~Kawasaki, M.~Yamaguchi, T.~Yanagida,
  Phys.\ Rev.\ D {\bf 63}, 103514 (2001)
  [hep-ph/0011104].
  
\bibitem{Kallosh:2010ug} 
  R.~Kallosh, A.~Linde,
  JCAP {\bf 1011}, 011 (2010)
  [arXiv:1008.3375 [hep-th]].
  
\bibitem{Kallosh:2010xz} 
  R.~Kallosh, A.~Linde, T.~Rube,
  Phys.\ Rev.\ D {\bf 83}, 043507 (2011)
  [arXiv:1011.5945 [hep-th]].

\bibitem{Takahashi:2010ky} 
  F.~Takahashi,
  Phys.\ Lett.\ B {\bf 693}, 140 (2010)
  [arXiv:1006.2801 [hep-ph]].
  
\bibitem{Nakayama:2010kt} 
  K.~Nakayama, F.~Takahashi,
  JCAP {\bf 1011}, 009 (2010)
  [arXiv:1008.2956 [hep-ph]];
  JCAP {\bf 1102}, 010 (2011)
  [arXiv:1008.4457 [hep-ph]];
  JCAP {\bf 1011}, 039 (2010)
  [arXiv:1009.3399 [hep-ph]].

\bibitem{Harigaya:2012pg} 
  K.~Harigaya, M.~Ibe, K.~Schmitz, T.~T.~Yanagida,
  arXiv:1211.6241 [hep-ph].
  
\bibitem{Silverstein:2008sg} 
  E.~Silverstein, A.~Westphal,
  Phys.\ Rev.\ D {\bf 78}, 106003 (2008)
  [arXiv:0803.3085 [hep-th]].
    
\bibitem{McAllister:2008hb} 
  L.~McAllister, E.~Silverstein, A.~Westphal,
  Phys.\ Rev.\ D {\bf 82}, 046003 (2010)
  [arXiv:0808.0706 [hep-th]].
  
\bibitem{Peiris:2013opa} 
  H.~Peiris, R.~Easther, R.~Flauger,
  arXiv:1303.2616 [astro-ph.CO].
  
\bibitem{Destri:2007pv} 
  C.~Destri, H.~J.~de Vega, N.~G.~Sanchez,
  Phys.\ Rev.\ D {\bf 77}, 043509 (2008)
  [astro-ph/0703417].
  
\bibitem{Croon:2013ana} 
  D.~Croon, J.~Ellis, N.~E.~Mavromatos,
  arXiv:1303.6253 [astro-ph.CO].
  
\bibitem{Cremmer:1983bf}
  E.~Cremmer, S.~Ferrara, C.~Kounnas and D.~V.~Nanopoulos,
  Phys.\ Lett.\ B {\bf 133}, 61 (1983).

\bibitem{Ellis:1983sf}
  J.~R.~Ellis, A.~B.~Lahanas, D.~V.~Nanopoulos and K.~Tamvakis,
  Phys.\ Lett.\ B {\bf 134}, 429 (1984);
  J.~R.~Ellis, C.~Kounnas and D.~V.~Nanopoulos,
  Nucl.\ Phys.\ B {\bf 247}, 373 (1984);
  J.~R.~Ellis, C.~Kounnas and D.~V.~Nanopoulos,
  Nucl.\ Phys.\ B {\bf 241}, 406 (1984).
  
\bibitem{Murayama:1993xu} 
  H.~Murayama, H.~Suzuki, T.~Yanagida, J.~'i.~Yokoyama,
  Phys.\ Rev.\ D {\bf 50}, 2356 (1994)
  [hep-ph/9311326].
  
  
  \bibitem{Linde:1983fq}   
  A.~D.~Linde, 
  Pisma Zh.\ Eksp.\ Teor.\ Fiz.\  {\bf 37}, 606 (1983)
  [JETP Lett.\  {\bf 37}, 724 (1983)];
  Phys.\ Lett.\ B {\bf 132}, 317 (1983).

\bibitem{Kallosh:2007wm} 
  R.~Kallosh, A.~D.~Linde,
  JCAP {\bf 0704}, 017 (2007)
  [arXiv:0704.0647 [hep-th]].
  
  
  
\bibitem{Liddle:2000cg} 
  A.~R.~Liddle and D.~H.~Lyth,
  ``Cosmological inflation and large scale structure,''
  Cambridge, UK: Univ. Pr. (2000).
  
  \bibitem{Fukugita:1986hr}
  M.~Fukugita, T.~Yanagida,
  Phys.\ Lett.\  {\bf B174}, 45 (1986).

  
\bibitem{Kawasaki:2006gs} 
  M.~Kawasaki, F.~Takahashi and T.~T.~Yanagida,
  Phys.\ Lett.\ B {\bf 638}, 8 (2006)
  [hep-ph/0603265];
  Phys.\ Rev.\ D {\bf 74}, 043519 (2006)
  [hep-ph/0605297];
  T.~Asaka, S.~Nakamura and M.~Yamaguchi,
  Phys.\ Rev.\ D {\bf 74}, 023520 (2006)
  [hep-ph/0604132];
  M.~Dine, R.~Kitano, A.~Morisse and Y.~Shirman,
  Phys.\ Rev.\ D {\bf 73}, 123518 (2006)
  [hep-ph/0604140];
  M.~Endo, K.~Hamaguchi and F.~Takahashi,
  Phys.\ Rev.\ D {\bf 74}, 023531 (2006)
  [hep-ph/0605091].
  
\bibitem{Endo:2006qk} 
  M.~Endo, M.~Kawasaki, F.~Takahashi, T.~T.~Yanagida,
  Phys.\ Lett.\ B {\bf 642}, 518 (2006)
  [hep-ph/0607170].
  
\bibitem{Endo:2007ih} 
  M.~Endo, F.~Takahashi and T.~T.~Yanagida,
  Phys.\ Lett.\ B {\bf 658}, 236 (2008)
  [hep-ph/0701042].
  
\bibitem{Endo:2007sz} 
  M.~Endo, F.~Takahashi and T.~T.~Yanagida,
  Phys.\ Rev.\ D {\bf 76}, 083509 (2007)
  [arXiv:0706.0986 [hep-ph]].

\bibitem{Nakayama:2012hy} 
  K.~Nakayama, F.~Takahashi and T.~T.~Yanagida,
  Phys.\ Lett.\ B {\bf 718}, 526 (2012)
  [arXiv:1209.2583 [hep-ph]].



\end{thebibliography}
\end{document}